\documentclass[twocolumn,showpacs,pre,preprintnumbers,floatfix]{revtex4}

\usepackage{listings}
\usepackage{amssymb}
\usepackage{amsmath}
\usepackage{epsf}
\usepackage{epsfig}

\usepackage{url}
\usepackage{subfigure}
\usepackage{setspace}
\usepackage{lineno}
\usepackage{mathrsfs}
\usepackage{soul}
\usepackage{color}
\usepackage{ulem}

\usepackage{graphicx}
\usepackage{pict2e}

\newcommand{\vs}{{\bf s}}

\newcommand{\bc}{\begin{center}}
\newcommand{\ec}{\end{center}}

\newcommand{\be}{\begin{equation}}
\newcommand{\ee}{\end{equation}}
\newcommand{\bea}{\begin{eqnarray}}
\newcommand{\eea}{\end{eqnarray}}

\begin{document}
\title{\bf Origins of Anomalous Transport in Disordered Media: Structural and Dynamic Controls}
\author{Yaniv Edery$^1$, Harvey Scher$^1$, Alberto Guadagnini$^2$ and Brian Berkowitz$^1$}
\affiliation{1. Department of Environmental Sciences and Energy
Research, Weizmann Institute of Science, 76100 Rehovot, Israel \\
2. (a) Dipartimento di Ingegneria Civile e Ambientale, Politecnico
di Milano, 20133 Milano, Piazza L. Da Vinci 32, Italy (b)
Department of Hydrology and Water Resources, University of
Arizona, Tucson, Arizona 85721, USA}

\begin{abstract}
We quantitatively identify the origin of anomalous transport in a
representative model of a heterogeneous system---tracer migration
in the complex flow patterns of a lognormally distributed
hydraulic conductivity ($K$) field. The transport, determined by a
particle tracking technique, is characterized by breakthrough
curves; the ensemble averaged curves document anomalous transport
in this system, which is entirely accounted for by a truncated
power-law distribution of local transition times $\psi(t)$ within
the framework of a continuous time random walk. Unique to this
study is the linking of $\psi(t)$ directly to the system
heterogeneity. We assess the statistics of the dominant preferred
pathways by forming a particle-visitation weighted histogram
$\{wK\}$. Converting the ln($K$) dependence of $\{wK\}$ into time
yields the equivalence of $\{wK\}$ and $\psi(t)$, and shows the
part of $\{wK\}$ that forms the power-law of $\psi(t)$, which is
the origin of anomalous transport. We also derive an expression
defining the power law exponent in terms of the $\{wK\}$
parameters. This equivalence is a remarkable result, particularly
given the correlated $K$-field, the complexity of the flow field
and the statistics of the particle transitions.

\end{abstract}

\maketitle

Anomalous transport was first discovered in the electron transport
of amorphous semiconductors \cite[]{SM75}. Such behavior is now
recognized to be ubiquitous in a broad range of physical
processes: chemical transport in porous and fractured geological
formations \cite[]{BS97,BCDS06,LBDC09,KDLBJ11,RDB12}, diffusion of
mRNA \cite[]{GC06} and chromosomal loci in live cells
\cite[]{WST10}, and tracer movement in micro-organism suspensions
\cite[]{LGGPG09}, crowded fluids \cite[]{SW09}, gels
\cite[]{KDM05} and porous media as affected by biofilm growth
\cite[]{SGCG04}. Anomalous transport in disordered media is
manifested by nonlinear time dependence of the particle
displacement mean, ${\bar\ell} \propto t^{\beta}$, and
${\bar\ell}/$rms $=const$ for $\beta <1$ (with rms the root mean
square particle displacement), as well as by long tails in the
spatial and temporal distributions of particles.

\par
As noted in many of these studies, the mechanisms for anomalous
transport and diffusion are varied, e.g., flow in heterogeneous
fracture networks and porous media, multiple trapping, chemical
sorption, migration in crowded environments and electron transfer
(hopping), and yet the time dependence is the same. In this
Letter, the origins of anomalous transport are identified through
an analysis of the particle pathway statistics in a completely
controllable model disordered geophysical system. We show the
transport cannot be explained solely by the structural knowledge
of the disordered medium. We probe the nuanced dynamic processes
and find that the basic determinant of the distribution of local
transition times, which defines the anomalous transport, is a
conductivity histogram weighted by the particle flux. This
remarkable relationship is quantitatively exact.

\par
The system is a geophysical model, based on a lognormally
distributed and spatially correlated hydraulic conductivity ($K$)
random field in a fluid saturated, two-dimensional (2D) domain.
The field is characterized by a variance of ln($K$), $\sigma^2$.
We determine the steady-state uniform (in the mean) flow and the
transport of tracer particles through it as a function of head
gradients and $\sigma^2$. The transport, determined by a particle
tracking technique, is characterized by breakthrough curves
(BTCs), quantifying the arrival of an injected plume of particles
at the domain outlet. The ensemble averaged BTC for this
disordered system documents anomalous (or ``non-Fickian'')
transport. Other studies have attempted to quantify the effect of
the underlying heterogeneous structure of the $K$ distribution on
BTC patterns; they note the occurrence of preferential flow paths
and long BTC tails, but quantification of the relationship between
$K$ field statistics and transport parameters remains obscure
\cite[e.g.,][]{WCS08,BZWTLG11}. Here, not only is it shown that
anomalous behavior is accounted for within the theoretical
framework of a continuous time random walk (CTRW), but an entirely
new assessment---the statistical analysis of flux-weighted
particle pathways---is linked directly to the CTRW framework.

\par
The discrete (in space), temporal semi-Markov CTRW transport
equations leads to the (Laplace space) working continuum transport
equation for the normalized concentration $c(\vs,t)$ for an
ensemble-averaged system:
\begin{equation} \label{eq:Ltrans}
      u\tilde{c}({\bf s},u) - c_0({\bf s}) = -\tilde{M}(u)
      [ {\mathbf v}_{\psi} \cdot \nabla \tilde{c}({\bf s},u) -
      {\mathbf D}_{\psi} : \nabla\nabla \tilde{c}({\bf s},u)]
\end{equation}
where $\tilde{M}(u) \equiv {\bar t}u \tilde{\psi}(u) /
[1-\tilde{\psi}(u)]$ is a memory function, with the Laplace
transform of a function $f(t)$ denoted by $\tilde{f}(u)$, ${\bar
t}$ is a characteristic time, $\psi(t)$ is the probability rate
for a transition time $t$ between sites, and ${\bar t}{\bf
v}_{\psi}$ and ${\bar t}{\bf D}_{\psi}$ are, respectively, the
first and second moments of $p({\bf s})$, the probability
distribution of the length of the transitions. The ``transport
velocity'', ${\bf v}_{\psi}$, is distinct from the ``average fluid
velocity'', ${\bf v}$, due to the heterogeneity.

\par
Unique to this study is the linking of $\psi(t)$ directly to the
heterogeneity of our model system. The key feature of the CTRW
distribution $\psi(t)$ that we use, which has been successful in
analyzing a number of laboratory and field observations, is a
truncated power law (TPL) distribution of the site-to-site
transition times with an evolution to Fickian behavior:
\begin{equation}\label{TPL}
\psi(t) = \frac{n}{t_1}\exp(-t/t_2)/(1+t/t_1)^{1+\beta}
\end{equation}
where $n \equiv (t_1/t_2)^{-\beta} \exp(- t_1/t_2 )/\Gamma(-\beta,
t_1/t_2 )$ is a normalization factor (for large $t_2$, $n \approx
\beta$ \cite[]{BES08}), $\beta$ is a measure of the transition
time spectrum, $t_1$ ($= {\bar t}$ in \eqref{eq:Ltrans}) is a
characteristic time, e.g., for median transitions between sites,
$t_{2}$ is a ``cutoff'' time, and $\Gamma (a, x)$ is the
incomplete Gamma function \cite[]{Abra}. For transition times $t_1
< t < t_2$, $\psi(t)$ behaves as a power law $\propto (t/t_1)^{-1
- \beta}$ while for $t
> t_2$, $\psi(t)$ decreases exponentially; thus a finite $t_{2}$
enables evolution from non-Fickian to Fickian transport. A pure
exponential $\psi(t)=\lambda \exp(-\lambda t)$ reduces the CTRW
transport equation \eqref{eq:Ltrans} to the advection-dispersion
equation (ADE): ${{\partial c ({\bf s},t)}/{\partial t}} = - {\bf
v}({\bf s}) \cdot\nabla c({\bf s},t) + \nabla \cdot\nabla ({\bf
D}({\bf s})c({\bf s}, t))$ where ${\bf v}(\bf s)$ is the velocity
field and ${\bf D}({\bf s})$ is the dispersion tensor. The basic
question now is: what aspects of the interplay between the
detailed particle dynamics and the structural features of the
$K$-field give rise to \eqref{TPL}?

\par
\begin{figure}
\noindent 
\includegraphics[scale = 0.4]{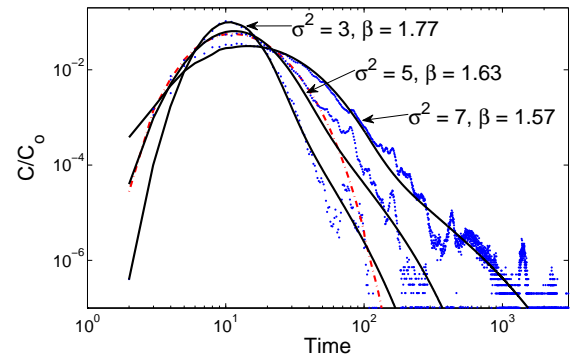}
\caption{(Color online) Ensemble breakthrough curves
(points) for three ln($K$) variances ($\sigma^2 = 3, 5, 7$) and a
head gradient of 100, and corresponding CTRW fits (curves), with
values of $v_{\psi}$, $D_{\psi}$ $\beta$, $t_1$, $t_2$ in
\eqref{TPL} of, respectively, [6.0, 15.8, 1.77, 0.055,
$10^{1.6}$], [5.9, 30, 1.63, 0.08, $10^{2.0}$], [3.8, 60, 1.57,
0.063, $10^{3.0}$]. Also shown is a fit of the ADE (dashed-dotted
curve, for $\sigma^2 = 5$ with $v = 3.4$, $D = 39$.  For
comparison, the average fluid velocity is 5.6. All values are in
consistent, arbitrary length and time units. The oscillations in
the BTC tails are caused by the formation of a limited set of
preferential channels (see Fig. \ref{Fig2} below), leading to
variations in the distribution of small numbers of particles
arriving at the outlet.} \label{Fig1}
\end{figure}

As background to answer this question, our method of model
construction and particle tracking is outlined in the
Supplementary Material (SM) \cite[]{SuppMat}.
The BTC is a key measure of the accumulative response of all the
transitions comprising the transport within the velocity field of
the domain. The ensemble mean BTCs (100 realizations) are seen in
Fig. \ref{Fig1}, with $\sigma^2 = 3, 5, 7$, and fits with 1D
solutions of the CTRW (using \eqref{TPL}) and the ADE. The
distinguishing feature is the broadness of the BTCs, which
increases with increasing $\sigma^2$. Overall, the CTRW
effectively captures the tails as well as the peaks of the BTCs.
The TPL values of $\beta$ show a clear trend to decrease (from
$1.77$ to $1.57$) with increasing $\sigma^2$ ($3$ to $7$), as
expected. This emphasizes the physical meaning of $\beta$: it is a
generalized dispersion parameter that captures the entire shape of
the BTC and not only a width of a normal curve (i.e., $\bf{D}$).
%
The subtle interplay among the TPL parameters $\beta, t_1, t_2$,
which determine the shape of the entire TPL, is quantified in
terms of the particle paths (below).

\par
\begin{figure}
\noindent
\includegraphics[scale = 0.45]{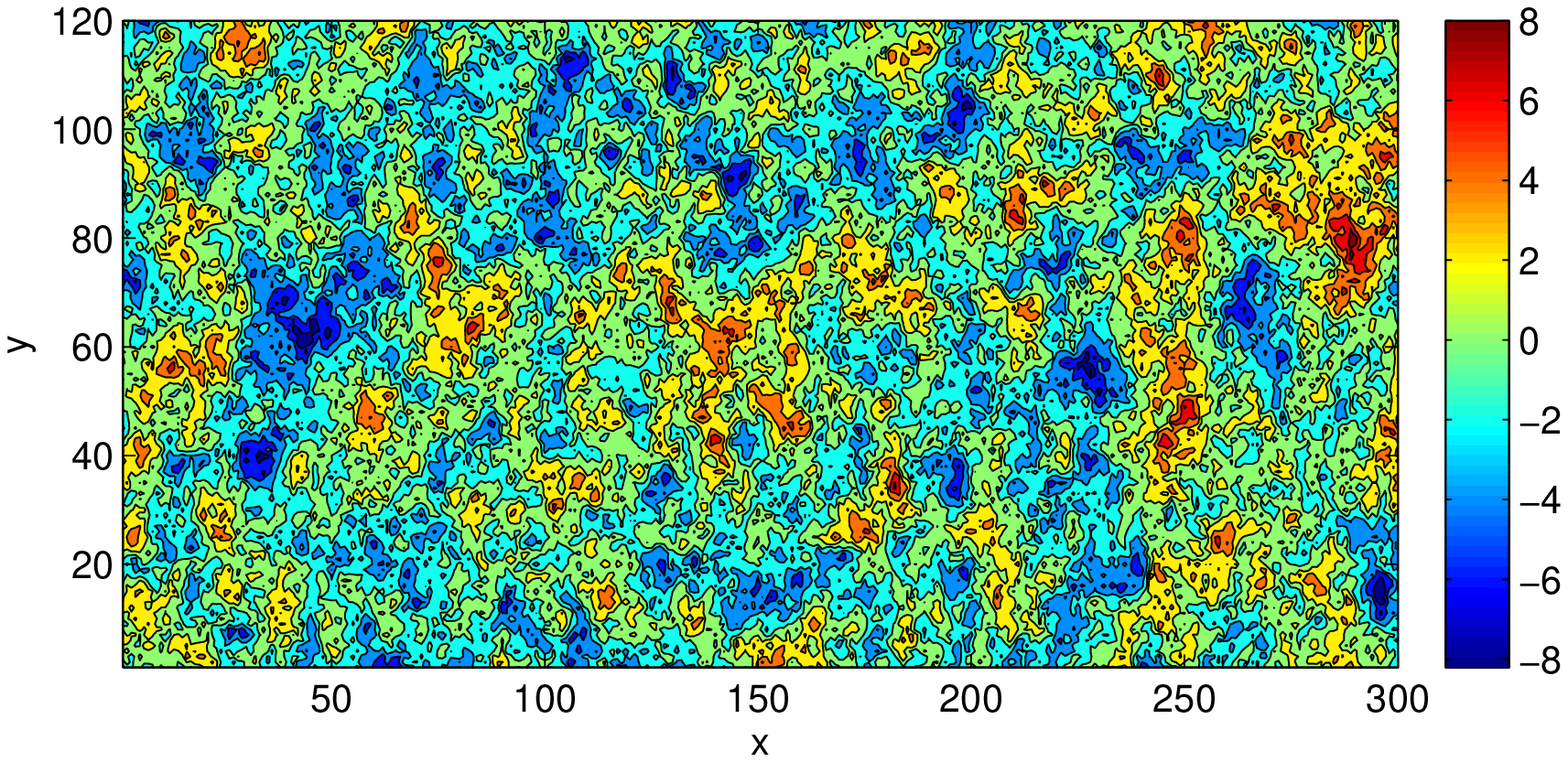}  \\ \vspace{2.2 mm}
\includegraphics[scale = 0.45]{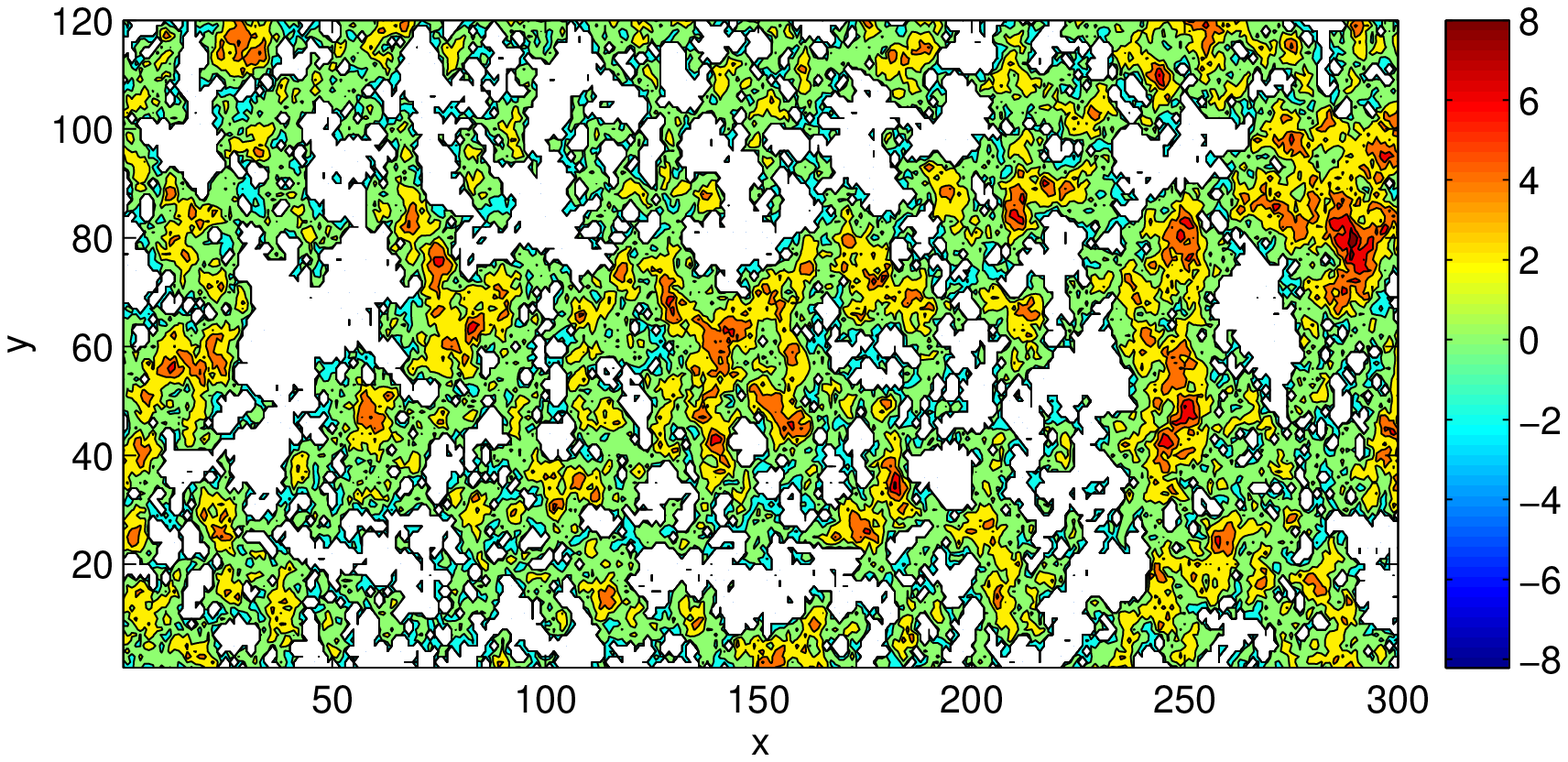}  \\ \vspace{2.2 mm}
\includegraphics[scale = 0.45]{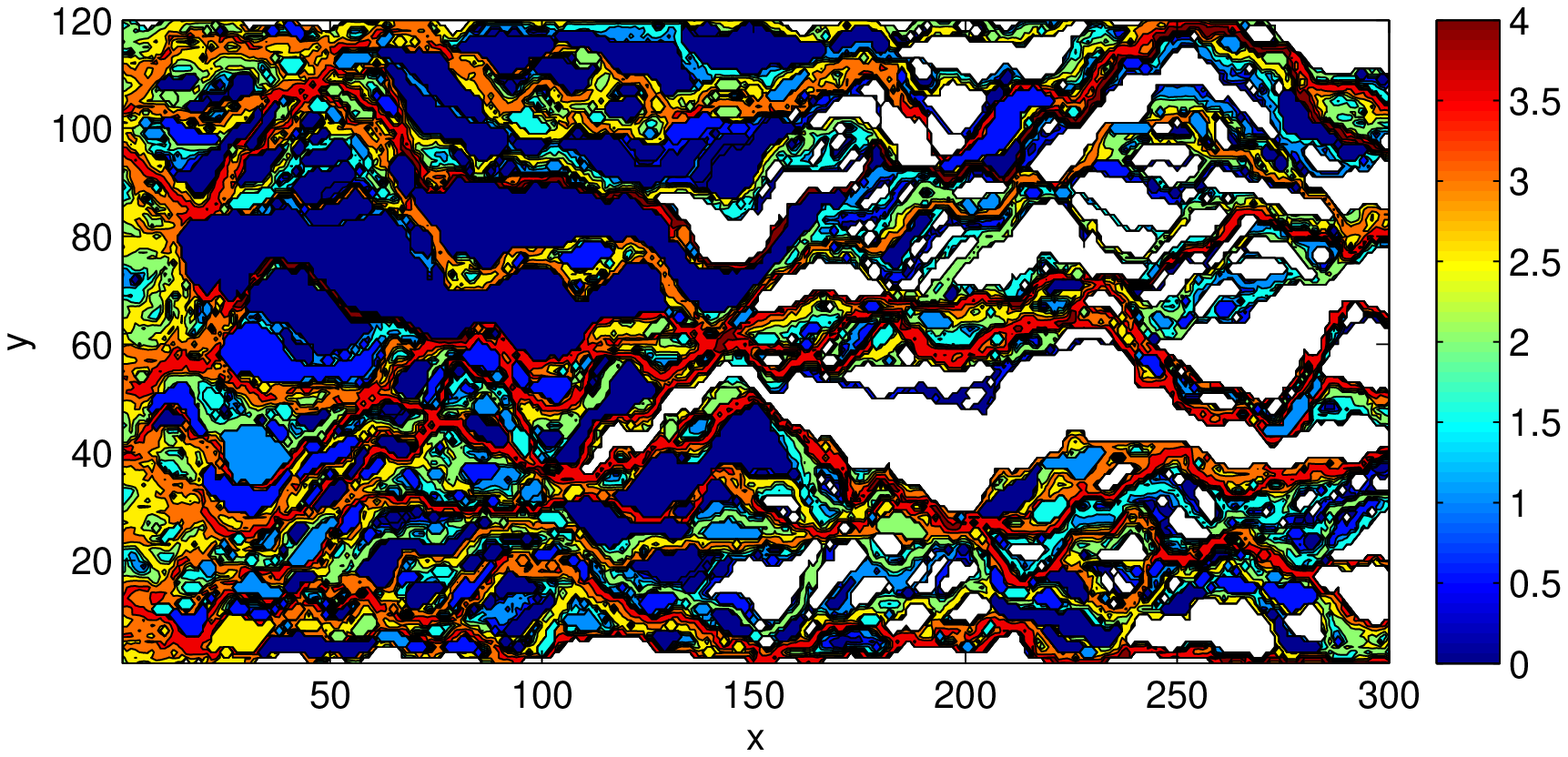}  \\ \vspace{2.2 mm}
\includegraphics[scale = 0.45]{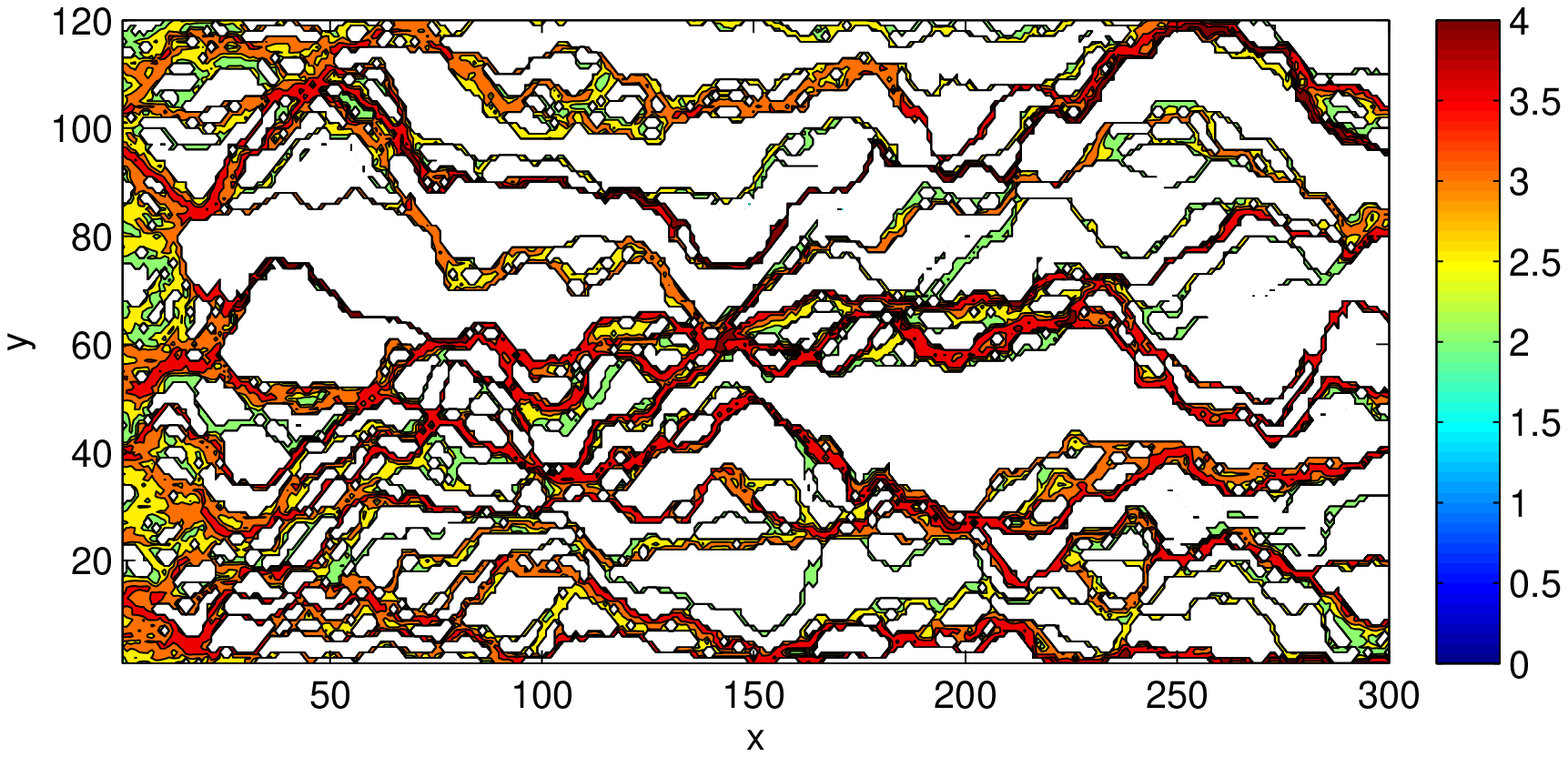}  \\ \vspace{2.2 mm}
\includegraphics[scale = 0.45]{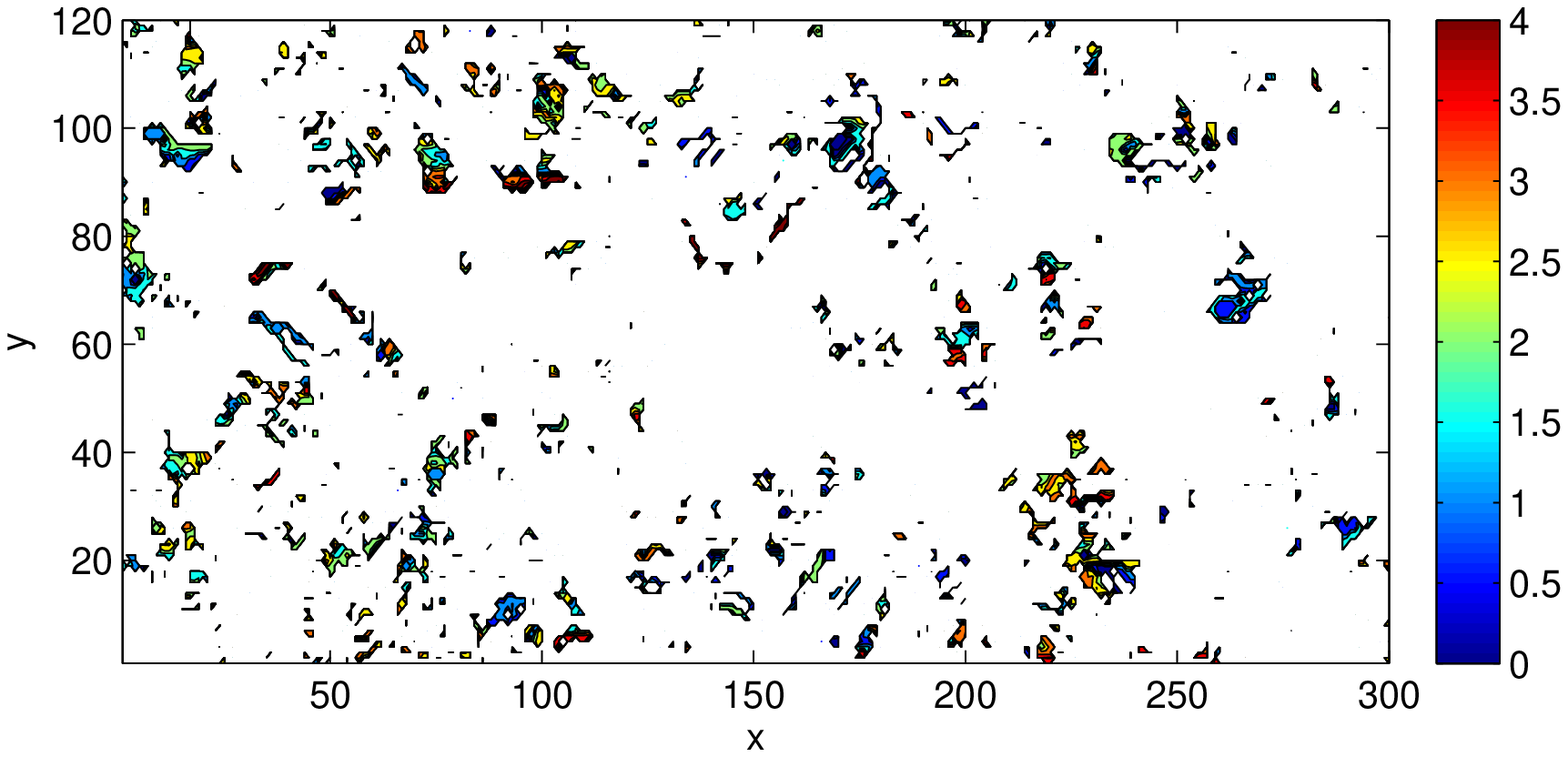}
\caption{(Color online) Spatial maps showing (a) full $K$ field,
(b) CPA (ln($K$) $< -1.13$), (c) particle paths, (d) preferential
particle paths, defined as paths through cells that each contain a
visitation of a minimum of 100 particles (= 0.1\% of the total
number of particles in the domain), (e) ``LCT jumps'' (see text).
Note that the color bars are in ln($K$) scale for (a)-(b), and log
number of particles for (c)-(e).} \label{Fig2}
\end{figure}

Fig. \ref{Fig2}a shows the heterogeneity of the $K$-field with
$\sigma^2 =5$. The mean ln($K$) is 0 with a 7 decade spread in $K$
over a statistically homogeneous map of a single realization. The
lowest values of ln($K$) tend to appear as local patches with a
concentric ring of moderately lower conductivity. We investigate
how this map manifests the preferred particle pathways. As a
reference, Fig. \ref{Fig2}b shows a path constructed by excluding
cells where the conductivity is lower than a given threshold; the
threshold value is lowered iteratively until a connected
(percolation) path is formed. This type of critical path analysis
(CPA) \cite[]{AHL71,Kirk71} has been linked, using percolation
theory scaling arguments, to anomalous transport and CTRW theory
\cite[]{HSEG11,GSH12,Sah12}. In Fig. \ref{Fig2}c, we superimpose
on the full conductivity field (Fig. \ref{Fig2}a) the number of
particles visiting each cell. The striking feature that emerges
is the occurrence of preferential particle paths
which are so dominant that the difference in particle
visitation in various cells ranges from zero to 10\% of the total
number of particles in the simulation. The white areas where
particles do not enter
have an effect on the surrounding areas, confining the
preferential paths to converge between low conductivity areas.

\par
Figure \ref{Fig2}d shows a sparser set of preferential paths than
Fig. \ref{Fig2}c, generated by recording only those cells having
visitations of $\ge$100 particles (0.1\% of all particles).
The color contrasts show an admixture of the higher conductivities
in the paths, however low conductivity cells are still present. We
find (not shown) that as residence time increases (i.e., for
smaller head gradients), the overall transport evolves to a biased
Gaussian (denoting Fickian) behavior for time scales $>t_2$, with
a BTC described by the ADE (see Fig. \ref{Fig1}). Significantly,
though, the particle flux in this limit is not spatially uniform
across the domain cross sections, as commonly envisaged for
application of the ADE; rather, the flux is still, largely, in
limited preferential pathways. It is illuminating to observe
deviations from the particle pathways defined by the CPA. Fig.
\ref{Fig2}e shows the ``lower conductivity transitions'' (LCTs)
--- defined for convenience as cells with $K$ values less than the
CPA threshold --- taken from the paths in Fig. \ref{Fig2}d.
Clearly, the critical path is insufficient to predict or estimate
the actual particle movement. Indeed, CPA and percolation scaling
arguments, based entirely on the $K$ field structure
\cite[]{Sah12}, do not include the significant influence of the
transitions below the threshold and residence time effects (see
below).

\par
Two histograms are shown in Fig. \ref{Fig3}, together with a
quantitative measure of the LCTs. One histogram corresponds to the
conductivity field in Fig. \ref{Fig2}a, while the other derives
from the preferred paths of Fig. \ref{Fig2}c weighted by the
particle visitation in each cell, which we designate $\{wK\}$. The
mean and skewness of $\{wK\}$ are significantly larger than those
for the full $K$ field (Fig. \ref{Fig3} caption). The mean of
$\{wK\}$ is a quantitative measure of the particle selectivity of
the higher conductivity cells. The fraction of the weighted
conductivities considered as LCT is 11.5\%. Hence these LCTs are
significant, and as we show below, are responsible for the long
tail in the BTC. The $\{wK\}$ in Fig. \ref{Fig3} is for one
realization; averaging the weighted mean over 100 realizations for
the same $\sigma^2$
follows this pattern closely. Moreover, similar behavior was
observed for a variety of realizations with different ln($K$)
variances (see below).

\par
The $\{wK\}$ is the basic characterization of transport in our
model. This is seen clearly by converting the ln($K$) axis to time
$t$. Based on Fig. \ref{Fig3}, for each $K$ bin, we determined the
head gradient over each associated cell, and obtained an average
head gradient (weighted by the relative number of particles
passing through these cells). We then determined the average
residence time in these cells, using Darcy's law for flow, $\Delta
t = \theta (\Delta x)^2/(K \Delta h)$, where $\theta$ is porosity
($=0.3$) and $\Delta h$ is the weighted average head difference
over the cell. Determining these average times for cells in all
$K$ bins, we obtain a frequency (weighted by the number of
particles) of particle residence times in all cells in the domain.
Dividing by $\Delta t$ to obtain equal bin sizes, yields the
probability density result for an ensemble of 100 realizations is
shown in Fig. \ref{Fig4}; the entire density is coincident with
the TPL \eqref{TPL} using $\beta=1.63$. Significantly, the
$\beta$, $t_1$ and $t_2$ values of \eqref{TPL} used here are the
same as those used to determine the BTC in Fig. \ref{Fig1}
($\sigma^2=5$). The statistical analysis of particle paths, which
renders the probability density $\{wK\}$, leads directly to the
CTRW framework of the probability density $\psi(t)$; indeed, they
are the same. The red portion of the curve corresponds to the
power-law region; by equating the logarithmic derivatives of both
curves we can develop an analytic expression for $\beta$ in terms
of $\{wK\}$ parameters. We first write $K=C/t$ (following Darcy's
law), the time associated with each $K$ being the transit time
across a ``$K$-bin''. The points in Fig. \ref{Fig4} are determined
from the numerical $\{wK\}$ for each realization. The mean of the
data in Fig. \ref{Fig4} is matched by $f=n_k \exp[-(\ln K -
\mu)^2/(2 \sigma^2)]/t$, where $\mu$, the mean of $\{wK\}$, is a
function of $\sigma^2$ and the variance of $\{wK\}$ is
$\approx\sigma^2$ (confirmed numerically); $n_k$ is a
normalization constant. We compute the logarithmic derivative
$d\log f/d\log t$, and obtain (see SM \cite[]{SuppMat})
\begin{equation}\label{eq:betasigma}
\beta=(\mu - \ln K)/\sigma^2
\end{equation}
by equating it to -1-$\beta$, the log derivative of the TPL
\eqref{TPL} in the power law region.
%
This result has a slow time dependence for $\beta$; Fig.
\ref{Fig4} has some curvature in the power-law region. The value
for $\beta$ is determined near the end of the range of small
ln($K$) (large time). Using representative ensemble values
($\sigma^2 = 5$, $\mu=1.5$) and choosing $\ln K = -6.5$ (the low
end (large time) of the LCTs in Fig. \ref{Fig3}) yields $\beta =
1.6$ (see Fig. \ref{Fig1}).

\par
A picture emerges from the path analysis of high particle flux
through the preferred paths, encountering relatively low
conductivities (aided by diffusion). The LCTs with highest
particle flux occur in or near the preferred paths.
The inclusion of these LCTs in the context of the preferred path
template is sufficient to provide to $\{wK\}$
a power law behavior of $\psi(t)$; this is the origin of
non-Fickian transport. This is highlighted in the power-law region
$t_1 < t < t_2$ of Fig. \ref{Fig4}; this region corresponds to a
ln($K$) range $\lesssim -1$ in the LCT region (red bars in Fig.
\ref{Fig3}). The role of the $K$ structure is to set up the flow
field, which forms the dynamical basis of the preferred paths. It
is the range of LCTs within this context that accounts for the
anomalous transport, as demonstrated here with the quantitative
relation $\{wK\} \leftrightarrow \psi(t)$.


\par
In conclusion, we have quantitatively identified the origin of
anomalous transport in a richly representative model of a
heterogeneous system. We proceeded via two levels. In the first,
we determined the BTCs, which are the cumulative particle arrival
times (first passage times). The BTCs document the anomalous
nature of the transport by the power-law dependence of late times
tails. The BTCs were fit with the solutions of the ADE and the
CTRW transport equation \eqref{eq:Ltrans} using a TPL \eqref{TPL}
form of $\psi(t)$. The TPL accounts for the full shape of the
BTCs, with the TPL $\beta$ parameter decreasing with enhanced
disorder (larger $\sigma^2$) as expected. The TPL $\psi(t)$ hence
serves as a characterization of anomalous transport. In the second
level, we deeply examined the nature of the particle pathways
across the domain and established the dominance of preferred
pathways. We assessed the dynamic statistics of these paths by
forming a particle-visitation weighted histogram $\{wK\}$. We
showed that these paths were, mainly, linked high conductivity
cells with an important, sparse mix of a relatively small number
of LCTs. We converted the ln($K$) dependence of $\{wK\}$ into time
and demonstrated the equivalence of $\{wK\}$ and the TPL
$\psi(t)$. In effect, we show that one can derive the TPL
\eqref{TPL} directly from the statistical analysis of the actual
particle pathways! We further pinpointed the range of the LCTs as
corresponding to the power-law region of the TPL and derived a
simple expression \eqref{eq:betasigma} for $\beta$ in terms of
$\sigma^2$ and $\mu$. Thus all the information is contained in
$\{wK\}$---including all correlations in the transitions---and
directly yields the values of $\beta$ as well as $t_2$. This is a
remarkable result, particularly given the correlated $K$-field,
the complexity of the flow field and the statistics of the
particle transitions.

\par
This research was supported by the Israel Science Foundation
(Grant No. 221/11), and by the Israel Water Authority (Grant No.
450056884).


\newpage

\begin{figure}
\noindent
\includegraphics[scale = 0.4]{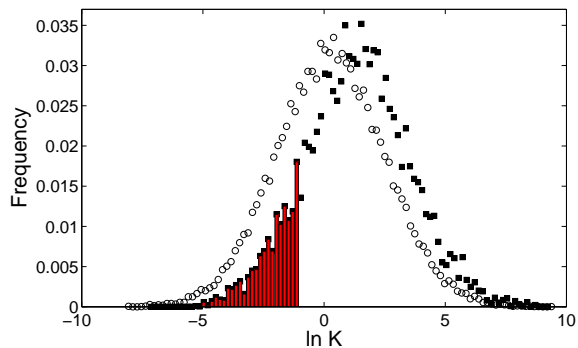}
\caption{(Color online) Conductivity histogram normalized by the
number of cells (open circles), corresponding to Fig. \ref{Fig2}a,
with mean ln($K$) of 0.26 and skewness of 0.03. Conductivity
histogram of the preferential particle paths (filled squares) (see
Fig. \ref{Fig2}c), weighted and normalized by the number of
particles visiting in each conductivity cell, $\{wK\}$; with
weighted mean of 1.43 and skewness is 3.89. Bars (denoted in red)
indicate the frequency of LCT (see Fig. \ref{Fig2}e) in the
weighted histogram of the preferential particle paths.}
\label{Fig3}
\end{figure}
\begin{figure}
\noindent
\includegraphics[scale = 0.4]{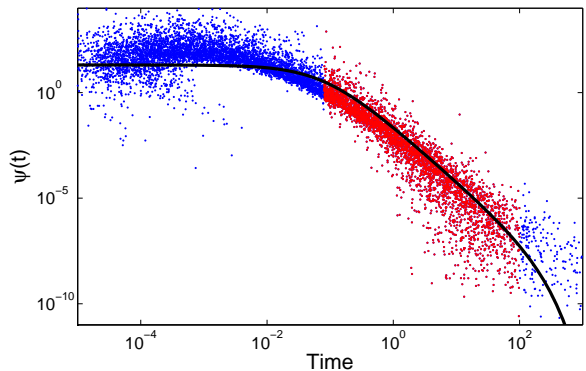}
\caption{(Color online) Ensemble particle-weighted conductivity
histogram ($\sigma^2=5$) for 100 realizations (points), based on
histograms such as shown in Fig. \ref{Fig3}, yielding a particle
transition time distribution within cells, representing $\psi(t)$
vs. $t$. The solid curve shows the TPL \eqref{TPL} with $\beta =
1.63$, $t_1 = 0.08$, $t_2 = 10^2$, identical to the values for the
fit shown in Fig. \ref{Fig1}. Highlighted (in red online) is the
power-law region $t_1 < t < t_2$ corresponding to $-7.5 < \ln(K) <
-1$. } \label{Fig4}
\end{figure}

\end{document}


\title{\bf Supplementary Material \\ Origins of Anomalous Transport in Disordered Media: Structural and Dynamic Controls}
\author{Yaniv Edery$^1$, Harvey Scher$^1$, Alberto Guadagnini$^2$ and Brian Berkowitz$^1$}
\affiliation{1. Department of Environmental Sciences and Energy
Research, Weizmann Institute of Science, 76100 Rehovot, Israel \\
2. (a) Dipartimento di Ingegneria Civile e Ambientale, Politecnico
di Milano, 20133 Milano, Piazza L. Da Vinci 32, Italy (b)
Department of Hydrology and Water Resources, University of
Arizona, Tucson, Arizona 85721, USA}

\begin{abstract}
We provide details of the model construction and particle tracking
methods, and the derivation of Eq. (4), employed in the main text.
\end{abstract}

\maketitle

\section{Model construction and particle tracking methods}
\par
Deterministic head values are prescribed along the left and right
boundaries of the 2D domain ($300 \times 120$ cells of uniform
length and width $\Delta = 0.2$; all quantities are given in
consistent units) while the top and bottom boundaries are
impervious. Values of ln($K$) in each cell are set equal to those
generated through a widely tested sequential Gaussian simulator
\cite{GJ93}. We generate different random realizations of
statistically homogeneous and isotropic, multivariate Gaussian
fields. Each field is associated with a spatial covariance of the
simple exponential type, with unit correlation length, $\cal L$ $=
1$, and ln($K$) variance, $\sigma^2$. Different values of
$\sigma^2$ are analyzed, to explore fields associated with mildly
to strongly heterogeneous conditions. In all scenarios, $\cal
L$/$\Delta = 5.0$, to accurately reproduce the adopted covariance
model for ln($K$) at all spatial lags of interest and its
influence on transport [e.g., \cite{AMGT89,RGS09}]. For each
generated ln($K$) field, we solve the corresponding deterministic
flow problem using Galerkin finite elements to obtain head and
velocity distributions.

\par
Solute movement in each realization is modeled by a Lagrangian
particle tracking method, to generate the BTCs. Particles are
injected, flux-weighted, along the left boundary. Starting from a
known location of the particles at time $t_k$, the displacement
vector ${\bf d}$ for each particle is given by the Langevin
equation
\begin{equation}\label{eq:disp}
{\bf d} = {\bf v}\left[{\bf x}({t}_k)\right] \delta{t} + \bf {d}_D
\end{equation}
with $\bf {v}$ the fluid velocity, $ \delta{t}$ the time step
magnitude, and $\bf {d}_D $ the diffusive displacement; $\xi$ is a
$N[0,1]$ random number and the modulus of $\bf {d}_D$ is $\xi
\sqrt{2 D \delta t}$.
The advective displacement in \eqref{eq:disp} is computed with the
velocity at $\bf x$. A uniform spatial step, $\delta{s}$, along
each particle trajectory is fixed. Preliminary numerical tests
confirmed that using $\delta{s} = \Delta /10$ did not introduce
numerical dispersion \cite[]{CK92}, and that consideration of
$10^5$ particles was sufficient. The magnitude of the time step
$\delta{t}$ in \eqref{eq:disp} is calculated as $\delta{t} =
\delta{s}/{v}$, $v$ being the modulus of $\bf {v}$. Reflection
boundary conditions are imposed at no-flow boundaries.

\section{Derivation of Equation (4)}
\par
The weighted histogram $\{wK \}$ is fit by a probability
distribution, which is then converted to time with $K_{ij}=\theta
\Delta x^2/\Delta h_{ij} \Delta t $, which we write as $K=C/t$
with the understanding that the time associated with each $K$ is
the transit time across a ``$K$-bin".

The mean of the data in Fig. 4 can be plotted as
\begin{equation} \label{f}
f=n_k\frac{\exp(-(\ln K-\mu)^2/2\sigma^2)}t
\end{equation}
where $\mu$, the mean of $\{wK\}$, is a function of $\sigma^2$ and
the variance of $\{wK\}$ is $\approx \sigma^2$; $n_k$ is a
normalization constant.

\par
We compute the logarithmic derivative of \eqref{f}, $d\log f/d\log
t$. First, we have

\begin{equation}
\log f = \log n_k+\frac{-(\ln K-\mu)^2}{2\sigma^2} \log(e) -\log t
.
\end{equation}

\par
Then
\begin{equation}
d\log f/d\log t = \frac{(\ln K-\mu)}{\sigma^2} \log(e) \ln(10) -1
\end{equation}
so that
\begin{equation}
d\log f/d\log t =\frac{(\ln K-\mu)}{\sigma^2} -1
\end{equation}
because
\begin{equation}
\log(e) \ln(10) = 1.
\end{equation}

\par
The above is equated to $-1-\beta$, the log derivative of the TPL
in the power law region, to obtain
\begin{equation}\label{beta}
\beta=\frac{(\mu-\ln K)}{\sigma^2}
\end{equation}
which is $>0$.

\par
The result in \eqref{beta} has a slow time dependence for $\beta$.
In Fig. 4 there is some curvature in the power-law region. The
value for $\beta$ is determined near the end of the range of small
$\ln(K)$ (large time).
